\documentclass[aps,prc,twocolumn,showpacs]{revtex4-1}

\usepackage{graphicx} 
\usepackage{epsfig} 
\usepackage{dcolumn} 
\usepackage{bm}         

\newcommand{\be}{\begin{equation}}
\newcommand{\ee}{\end{equation}}
\newcommand{\ba}{\begin{eqnarray}}
\newcommand{\ea}{\end{eqnarray}}
\newcommand{\bd}{\begin{displaymath}}
\newcommand{\ed}{\end{displaymath}}

\begin{document}

\title{Fluid Dynamical Prediction of Changed $v_1$-flow at LHC}

\author{
L.P.~Csernai$^{1,2,3}$,
V.K.~Magas$^{4}$,
H.~St\"ocker$^{3}$, and
D.D.~Strottman$^{1,3}$
} 
\affiliation{
$^1$ Institute of Physics and Technology, University of Bergen, Allegaten 55, 
5007 Bergen, Norway \\
$^2$ MTA-KFKI, Research Inst.\ of Particle and Nuclear Physics, 1525 Budapest, 
Hungary\\
$^3$ Frankfurt Institute for Advanced Studies - Goethe University, 
60438 Frankfurt am Main, Germany\\
$^4$ Departament d'Estructura i Constituents de la Mat\`eria, Universitat de 
Barcelona, 08028 Barcelona, Spain\\
}

\date{\today}

\begin{abstract}
Substantial collective flow is observed in collisions between Lead nuclei at 
LHC as evidenced by the  azimuthal correlations in the transverse 
momentum distributions of the produced particles.  
Our calculations indicate that the Global $v_1$-flow, which at RHIC peaked at negative 
rapidities (named as {\it 3rd flow component} or {\it anti-flow}), now at LHC is going to turn toward
forward rapidities (to the same side and direction as the projectile 
residue). 
Potentially this can provide a sensitive barometer to estimate the pressure and 
transport properties of the Quark-Gluon Plasma. 
Our calculations also take into account the initial state Center of Mass rapidity fluctuations, and 
demonstrate that these are crucial for $v_1$ simulations. 
In order to better study the transverse momentum flow dependence we suggest a new "symmetrized" 
$v_1^S(p_t)$  flow component; and we also propose a new method to disentangle  Global $v_1$ flow from the 
contribution generated by the random fluctuations in the initial state. 
This will enhance the possibilities of studying
the collective Global  $v_1$ flow both at the STAR Beam Energy Scan program
and at LHC. 
\end{abstract}



\maketitle

The first publication from the LHC heavy ion run presented 
amazingly strong elliptic flow, exceeding all measurements at lower
energies \cite{ALICE-Flow1}. 
This indicates strong equilibration and thermalization
at these energies in contrast to expectations of increasing transparency.
Just 6 month later ALICE has also measured the  $v_1$ flow \cite{QM2011}.

The overall picture indicated by the first $v_1$ results is very similar for RHIC and ALICE/LHC; 
namely that $v_1$ has three physical sources \cite{RHIC_V1,QM2011}:\\
  (i) the Global collective flow correlated with the reaction plane
      of the event {EP};\\
  (ii) the Random fluctuation flow of all $v_n$ varieties, where the 
      corresponding symmetry axes for (e.g. for $v_1$ and $v_3$) have no
      correlation with the reaction plane {EP}, instead they are observed with
      respect to a participant plane {PP} Event by Event (EbE) \cite{PP_method_QM,schenke}. The 
      participant planes are different for the neighbouring flow harmonics;\\
  (iii) at high momenta or high  pseudo-rapidity, ($1.5 \le |\eta| \le 4$), there
      are strong anti-flow peaks (in opposite direction with respect to classical bounce-off).  These appear only at 
      RHIC and LHC energies \cite{RHIC_V1,QM2011}, and there this is the strongest source of $v_1$. 
      These high momenta particles are not, and probably can not be, described by fluid dynamical models. It seems reasonable that they are generated in very early (pre-equilibrium) times of the reaction, and such an emission is anti-correlated with
      the projectile spectator in the reaction plane due to shadowing effect of the main reaction volume \cite{Ko,v1_cascade}. 
      The hybrid transport AMPT model provided a qualitative match to this $v_1$ flow component under special assumptions (switching off 'sting-melting') \cite{Ko}, what basically means very early (pre-equilibrium?) hadronization 
and freeze out in some parts of the reaction volume.

This article discusses the behaviour of the first (i), among these flow 
phenomena, which is the weakest at RHIC and LHC energies. We will 
also discuss, how to separate the Global $v_1$-flow, from the one produced
by random EbE fluctuations of the initial state (ii).

Collective flow is evidenced by the radial flow and, in non-central
collisions, by the asymmetric azimuthal distribution around the beam axis 
quantified by the functions $v_1(y,p_t), v_2(y,p_t), ...$ in the expansion
$$
\frac{d^3N}{dydp_td\phi} = \frac{1}{2\pi}\frac{d^2N}{dydp_t} \left[ 1 
+  2v_1 \cos(\phi) +  2v_2\cos(2\phi) + \cdot\cdot\cdot \  \right]
$$
where $y$ is the rapidity and $p_t$ is the transverse momentum and
$\phi$ is the azimuth angle in the transverse plane with respect to 
impact parameter vector, $\vec{b}$ .  

The observed large $v_2(p_t)$ has important consequences. As the
previously observed constituent quark number scaling indicates, the collective
flow must have developed in the Quark-Gluon Plasma (QGP) phase, and the flow
at the partonic level becomes observable after partons coalesce \cite{coalesce}.  
Theoretical calculations also indicate
that to explain the observed flow enhanced partonic interaction
is needed over perturbative QCD predictions \cite{Molnar}. Thus the QGP is
strongly interacting. At the same time theoretical estimates and
observations also indicate that the QGP is a nearly perfect fluid, with
minimal shear viscosity at the phase transition point 
\cite{Son,CKM}.

\begin{figure}
 \centering
 \includegraphics[angle=-90,width=3.5in]{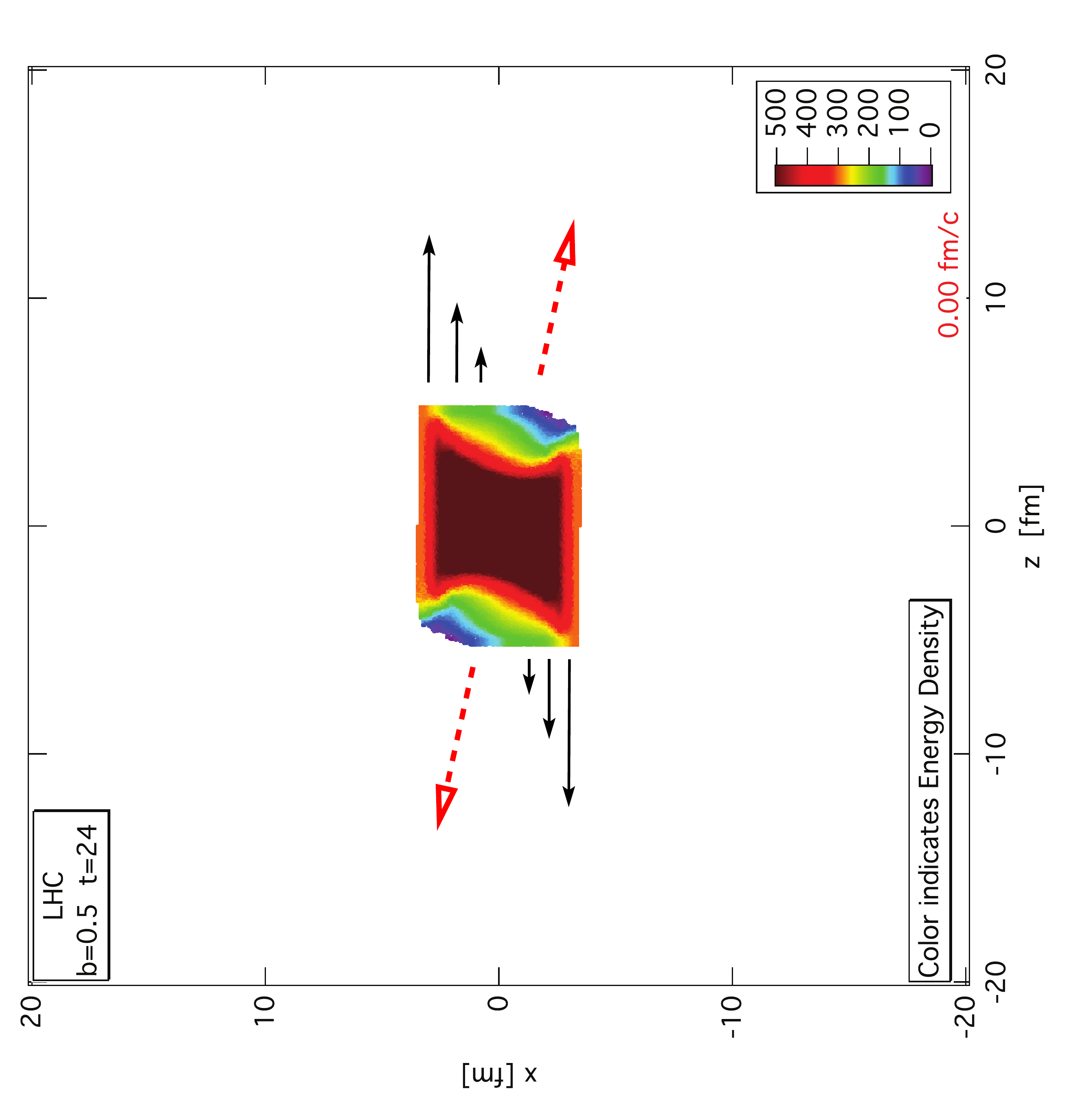}
\vskip -5mm
 \caption{(color online) 
Initial energy density [GeV/fm$^3$] 
distribution in the reaction plane, [x,y] 
for a Pb+Pb reaction at 1.38 + 1.38 A$\cdot$TeV collision energy
and impact parameter $b = 0.5 b_{max}$ at time 4 fm/c after the first 
touch of the colliding nuclei, this is when the hydro stage begins. 
The calculations are performed according to 
the effective string rope model  \cite{MCs001}. This tilted initial 
state has a flow velocity distribution, qualitatively shown by the arrows. 
The dashed arrows indicate the direction of the largest pressure gradient 
at this given moment.
}
\vskip -8mm
 \label{in-state}
\end{figure}
{\bf Our model.\ \ }
The energy-momentum tensor density for a perfect fluid  is $T^{\mu\nu} = 
(e+P)u^{\mu}u^{\nu}-Pg^{\mu\nu}$, where $P$ is the local pressure, 
$e$ is the local energy density, and $u^{\mu}=\gamma(1,\vec{v})$ is the 
local flow velocity.  We assume the MIT Bag Model
Equation of State during the whole calculation.

A fluid dynamical (FD) description of the nuclear matter is
considered here for Pb+Pb collisions at 1.38 + 1.38 A$\cdot$TeV.  
The matter expands until it reaches freeze-out (FO). 
The FD description does not constrain the FO: an external condition, 
for example a fixed FO temperature, is needed. The FD model we use
\cite{CC10,CC09,BC98}
can run well beyond the FO, so the location of physical FO can be 
selected afterwards as a space-time hyper-surface.

The (3+1)-dimensional, FD model \cite{CC10,CC09,BC98} uses the Particle in
Cell (PIC) method adapted to ultra-relativistic heavy ion collisions.
The numerical dissipation of the method was analyzed recently in
\cite{HSz}.
In this method, marker particles,  corresponding to fixed baryon 
charge, move in an  Eulerian grid. The calculation, describing the
 reaction, starts from an analytic initial state model 
\cite{MCs001},
based on longitudinally expanding strings
of the color-magnetic field.  The produced initial state, shown in 
Fig. \ref{in-state}, is tilted, and, thus, the direction of the largest 
pressure gradient is pointing in the "anti-flow" direction, what 
resulted in anti-flow peaks in simulations for RHIC and SPS 
\cite{CsR99,Baeu07}. However, one should not forget that this initial state 
also has a flow velocity distribution, which tends to further rotate it, 
i.e. effectively it has a large initial "angular momentum", what will change 
the direction of the strongest pressure gradient with time. 

\begin{figure}
 \centering
 \includegraphics[width=3.5in]{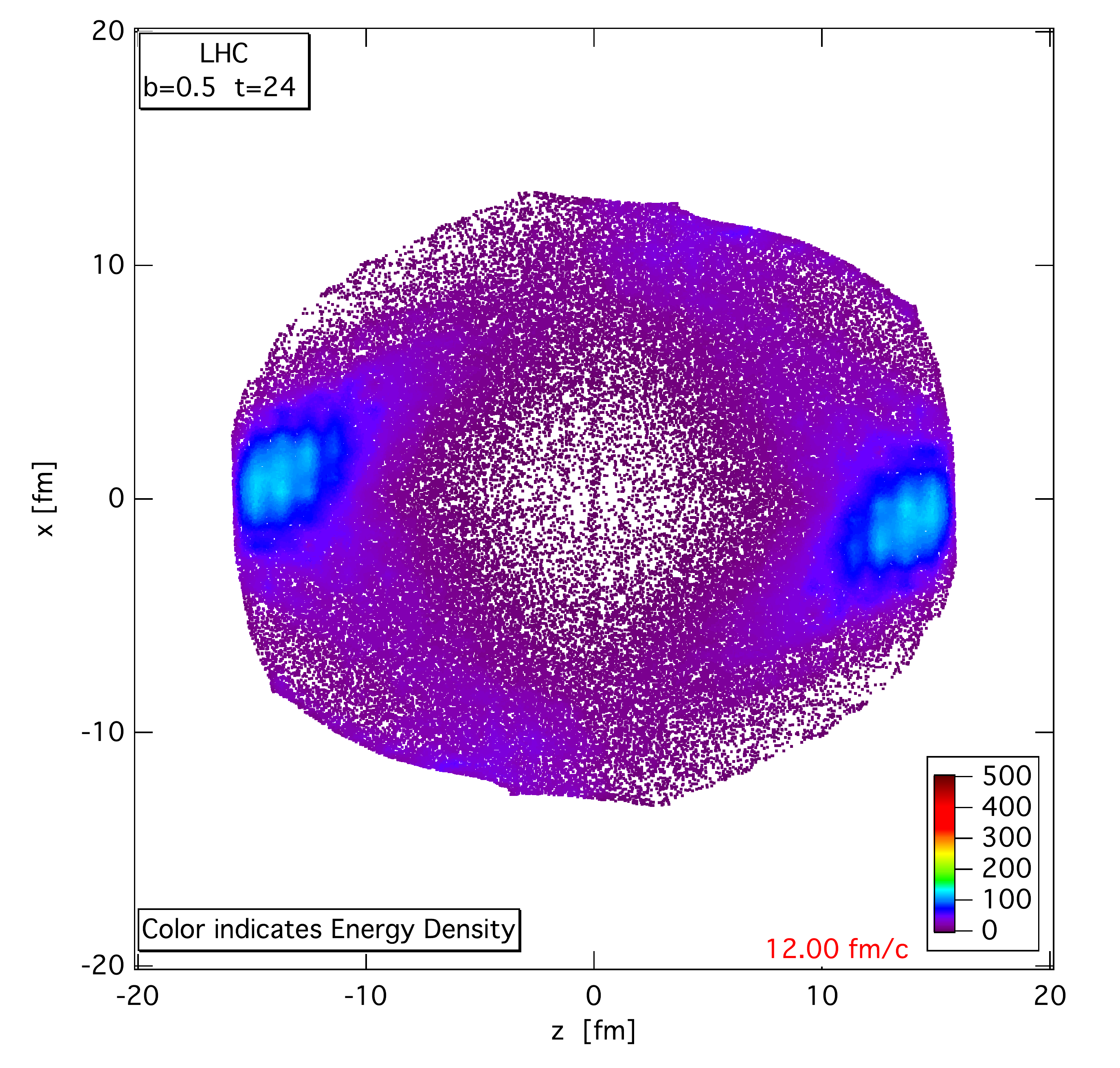}
\vskip -5mm
 \caption{(color online) The energy density [GeV/fm$^3$] 
distribution in the reaction plane, [x,z] 
for the reaction shown in Fig. \ref{in-state} at time 12 fm/c after 
the formation of the hydro initial state. 
The expected physical FO point is earlier
but this post FO configuration illustrates the
flow pattern. 
}
\vskip -5mm
\label{Fig0-Flow}
\end{figure}

Fig. \ref{Fig0-Flow} shows the energy density distribution in the reaction 
plane later. One may notice that the
final state is strongly rotated with respect to the initial one, due to the 
large initial "angular momentum", and the direction of strongest
transverse expansion points to $\Theta = 75/255^o$. Thus,  the upward moving 
matter is moving now forward  and the downward moving matter backward, in 
contrast to what happens at RHIC and SPS energies \cite{CC09}.

The fluid cells in the presented calculations were 
$(0.438 fm)^3$ for peripheral collisions, $b=0.5 - 0.7 b_{max}$.
While initially we had 2500- 5400 fluid cells containing matter, this
increased over 100 000 by the end of calculation. 
The higher energy at LHC results in a more explosive
expansion, which leads to an explosion shell with decreasing 
central density.  

In a simplest approach we assume a constant time FO
hypersurface.  Comparing measured multiplicity $b$-dependence at LHC 
with our FD multiplicity, we have chosen $t_{FO}=8$ fm/c after 
the formation of the hydro initial state.
The transition from pre FO QGP
to post FO ideal massless pion J\"uttner gas is calculated according to
the method described in ref. \cite{Yun10}, satisfying the
conservation laws. In this way for each fluid cell ${i}$, we obtain  
a flow velocity $\vec{v}^i=(\vec{v_t}^i,v_z^i)$ of the gas and 
its temperature $T^i$. 

Using the Cooper-Frye FO formula we obtain:
\begin{eqnarray}
v_n(y,p_t) & = & 
\frac{\sum_{i}^{cells}\;\int_0^{2\pi}
 \!\! d \phi\; f^i (y,\vec{p}_t)\; \cos n \phi}
{\sum_{i}^{cells}\;\int_0^{2\pi}\!\!  d \phi\; f^i (y,\vec{p}_t)}\,,
\label{vn}
\end{eqnarray}
where $f^i(y,\vec{p}_t)$ is the normalized momentum distribution for cell $i$; the angle $\phi$ is taken with respect to the
reaction plane. Then,
\begin{equation}
v_n(y)=
\frac{\sum_i^{cells} J_n(y,\vec{v}\,^i,T^i) cos(n\phi_0^i)}
{\sum_i^{cells} J_0(y,\vec{v}\,^i,T^i) }\,, 
\label{flow_y}
\end{equation}
\vspace{-4mm}
\begin{equation}
v_n(p_t)=
\frac
{\sum_i^{cells} B(\vec{v}\,^i,T^i,p_t) I_n(\gamma^i v_t^i p_t/T^i) 
cos(n\phi_0^i)}{\sum_i^{cells} B(\vec{v}\,^i,T^i,p_t) 
I_0(\gamma^i v_t^i p_t/T^i) }\,,
\label{flow_pt}
\end{equation}

\vspace{-4mm}
$$
J_n(y,\vec{v}\,^i,T^i)=\!
\int_0^{\infty}\!\!\! dp_t p_t^2 I_n(\gamma_t^i \tilde{v}_t^i p_t/T^i) 
e^{-\gamma_t^i p_t cosh(y-y_0^i)/T^i}
$$
\vspace{-4mm}
$$
B(\vec{v},T,p_t)=
e^{-\gamma p_t/T}\frac{1}{1-v_z^2}\left(v_z \frac{T}{\gamma}-p_t|v_z|\right)
$$
\vspace{-4mm}
$$
+\frac{p_t}{\sqrt{1-v_z^2}}
K_1\left(\gamma p_t \sqrt{1-v_z^2}/T,\gamma p_t/T \right)\,.
$$
where $y_0^i$ is the flow rapidity and $\phi_0^i$ is the azimuthal angle of 
the flow velocity in the transverse plane of the given cell ${i}$. 
In eq. (\ref{flow_y})
we have rewritten flow 4-velocity in the following way:
$u^\mu_i=\gamma_t^i(\cosh{y_0^i},\sinh{y_0^i},\vec{\tilde{v}_t^i})$, with 
$\vec{\tilde{v}_t^i}= \vec{v_t^i}/\sqrt{1-(v_z^i)^2}$,  
$\gamma_t^i=1/\sqrt{1-(\tilde{v}_t^i)^2}$.
 $I_n$ is 
a Bessel function, and 
$K_1(a,b)={1\over a} \int_b^{\infty}dx \sqrt{x^2-a^2}e^{-x}$ 
is a modified Bessel function of the second kind. 

\begin{figure}[h]
 \centering
 \includegraphics[width=3.3in]{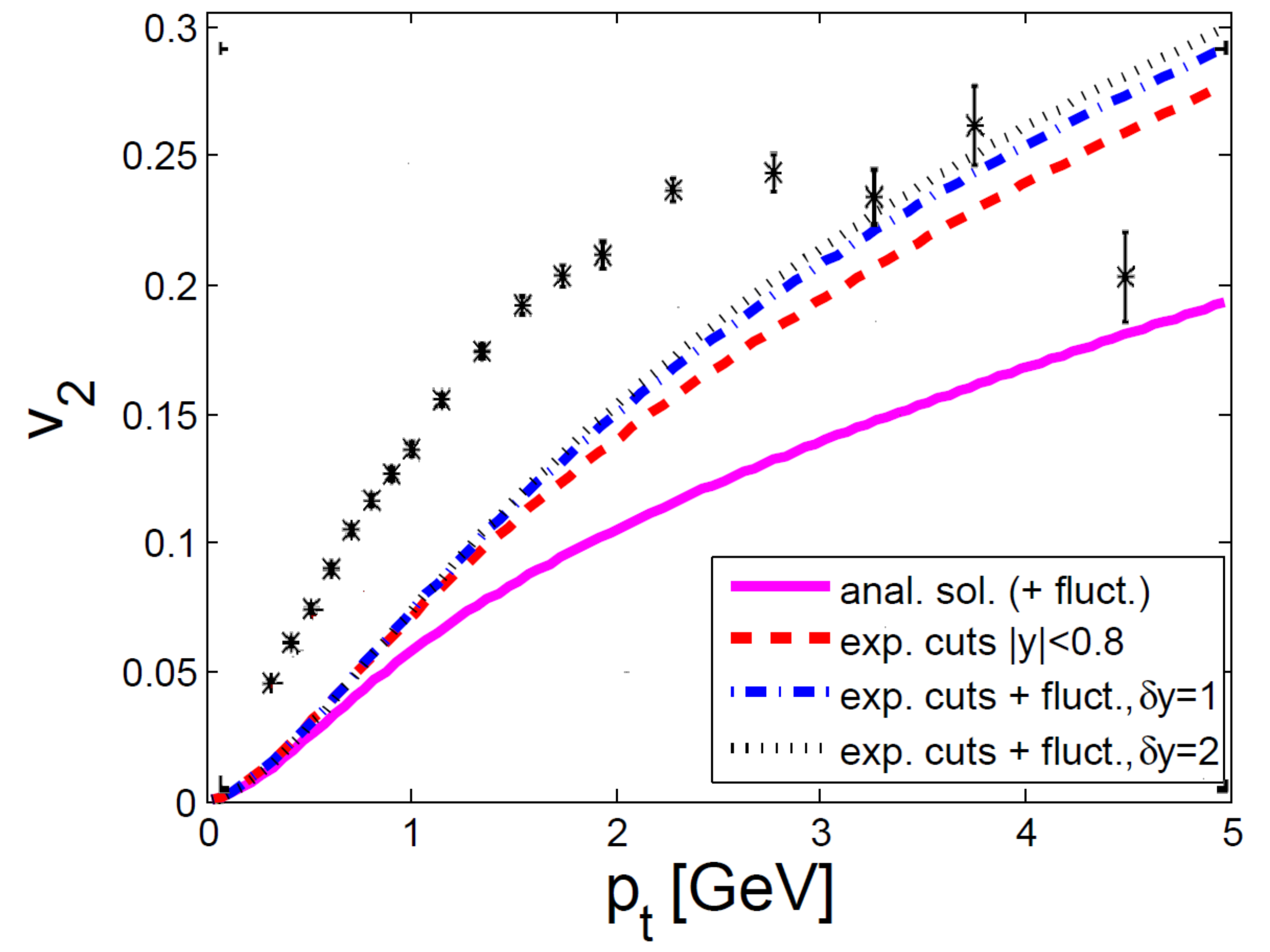}
\vskip -4mm
 \caption{The $v_2$ parameter calculated for ideal massless pion Juttner gas,
versus the transverse momentum, $p_t$ for $b = 0.7 b_{max}$, at
$t = 8$ fm/c  FO time. The magnitude of $v_2$ is comparable to the
observed $v_2$ at 40-50 \% centrality (black stars). 
See text for more explanation of different curves. }
\vskip -6mm
 \label{Fig4-v2pt}
\end{figure}

{\bf Results: $p_t$-dependence of the flow.\ \ }
The calculated $v_2(p_t)$ distributions are similar to the experimental
trends.
For illustration one calculated $v_2(p_t)$-distribution is presented 
in Fig. \ref{Fig4-v2pt}. 
The full curve, calculated according to eq. (\ref{flow_pt}),  is 
slightly below the experimental data. This can 
be attributed to the integral over the whole rapidity range, while 
the experiment is only for $| \eta | < 0.8 $, and to the initial state 
fluctuations, as discussed below.

\begin{figure}[h]
 \centering
 \includegraphics[width=3.3in]{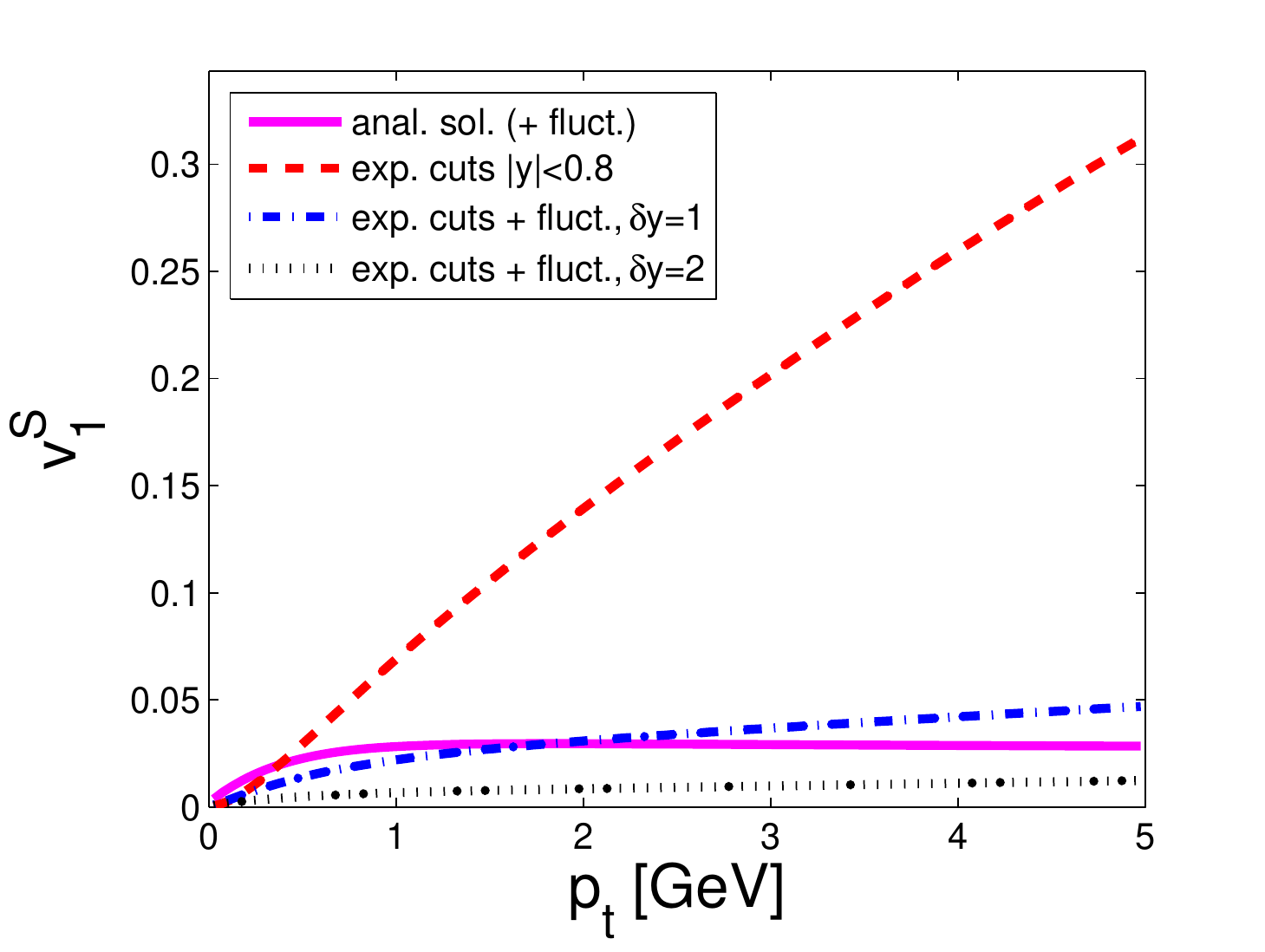}
\vskip -4mm
 \caption{(color online) 
The $v_1^S$ flow parameter calculated according to the eq. (\ref{V1S_pt}) 
for ideal massless pion Juttner gas,
versus the transverse momentum, $p_t$ for $b = 0.7 b_{max}$, at
$t = 8$ fm/c  FO time. See text for explanation of different curves.}
\vskip -4mm
\label{Fig6-v1spt-b5b7}
\end{figure}

As $v_1$ is an antisymmetric function of $y$, the $y$-integrated $v_1(p_t)$
value must vanish. In our calculation this is realized to an
accuracy better than $10^{-16}$. However, considering this obvious
asymmetry, we propose to constuct a symmetrized function $v_1^S$ reversing 
the $\vec{p}_t$-direction of backward going ($y<0$) particles:
\begin{eqnarray}
v_1^S(y,p_t) & = & 
\frac{ \sum_{i}^{cells}\; \int_0^{2\pi}
\!\!  d \phi\; f^i (y,{\rm sgn}(y)\cdot \vec{p}_t)\; \cos \phi}
{ \sum_{i}^{cells}\; \int_0^{2\pi} 
\!\! d \phi\; f^i (y,{\rm sgn}(y)\cdot\vec{p}_t)}\,,
\label{v1s}
\end{eqnarray}
where ${\rm sgn}(y)$ extracts the sign of rapidity. 
The idea stems from Danielewicz and 
Odyniecz \cite{DO85}.
In this way we get a non-vanishing $v_{1}^S(p_t)$ function, which will 
be also much less sensitive to the initial state fluctuations.
\begin{equation}
v_1^S(p_t)=
\frac{
\sum_i^{cells} 2D(\vec{v}\,^i,T^i,p_t) I_1(\gamma^i v_t^i p_t/T^i) 
cos(\phi_0^i)}
{\sum_i^{cells} B(\vec{v}\,^i,T^i,p_t) I_0(\gamma^i v_t^i p_t/T^i) }\,,
\label{V1S_pt}
\end{equation}
where
$
D(\vec{v},T,p_t)=e^{-\gamma p_t/T}\frac{v_z}{1-v_z^2}\frac{T}{\gamma}
$.
The  $v_{1}^{S}(p_t)$ parameter calculated in this way is shown
in Fig. \ref{Fig6-v1spt-b5b7}  (full line).

The ALICE team has made a detailed symmetry analysis of the low rapidity
component in the acceptance range of the ALICE TPC \cite{QM2011}. They introduced new quantities
\ba
v_1^{even/odd}(\eta,p_t) = [v_1(\eta,p_t) \pm v_1(-\eta,p_t)] /2\,,\\
v_1(\eta,p_t) = v_1^{even}(\eta,p_t)+v_1^{odd}(\eta,p_t)\,.
\ea
If we have a Global Mirror Asymmetric (MA)  $v_1$ flow only, as in our hydrodynamical simulations, 
then the $even$ component vanishes:
\ba v_1^{even}(\eta,p_t) = 0\,, \quad v_1^{odd}(\eta,p_t)=v_1(\eta,p_t)\,.\ea
A non-vanishing $v_1^{even}(\eta,p_t)$ can only come from the Mirror Symmetric (MS) part of random fluctuation flow.

Furthermore, if we will integrate over (pseudo-) rapidity, then, as it was discussed above, the Global $v_1(p_t)$ flow, correlated with the reaction plane of the event, is exactly zero. Thus, non-zero  $v_1^{even}(p_t)$ and $v_1^{odd}(p_t)$ can only come from the random fluctuation in the initial state. Without other assumption, we should not make any difference for MS and MA fluctuations, and thus these two components should be (approximately) equal: $v_1^{even}(p_t)=v_1^{odd}(p_t)$. 
The preliminary ALICE results \cite{QM2011} clearly comfirm the simple logical sequence. So, we can conlcude that $v_1^{even}(p_t)= v_1^{odd}(p_t)$ observed by ALICE collaboration come from the second $v_1$ flow source (ii), and can not tell us anything about Global $v_1$ flow (i). 
 
However, we can gain information about the $p_t$ dependence of the Global directed flow, if we repeat the same analysis, i.e. separation into {\it even} and {\it odd} components, for the  $v_{1}^S(y,p_t)$ function, introduced above, eq. (\ref{v1s}). Indeed, we obtain:
\vspace{-4mm}
\ba
v_1^{S,odd}(p_t)=v_{1,fluct.}^S(p_t)\,,\\
v_1^{S,even}(p_t) =v_1^S(p_t)+v_{1,fluct.}^S(p_t)\,,
\ea
where $v_{1,fluct.}^S(p_t)$ is a contribution to the $v_{1}^S$ function from the initial state fluctuations, which is approximately equal for both components, as discussed above. Thus, in this case we can separate the contributions from the (i) and (ii) sources:
\vspace{-4mm}
\ba
v_{1,fluct.}^S(p_t)=v_1^{S,odd}(p_t)\,,  \\
v_1^S(p_t)=v_1^{S,even}(p_t)-v_1^{S,odd}(p_t)\,.
\label{v1s_sep}
\ea

\begin{figure}[h]
 \centering
 \includegraphics[width=3.3in]{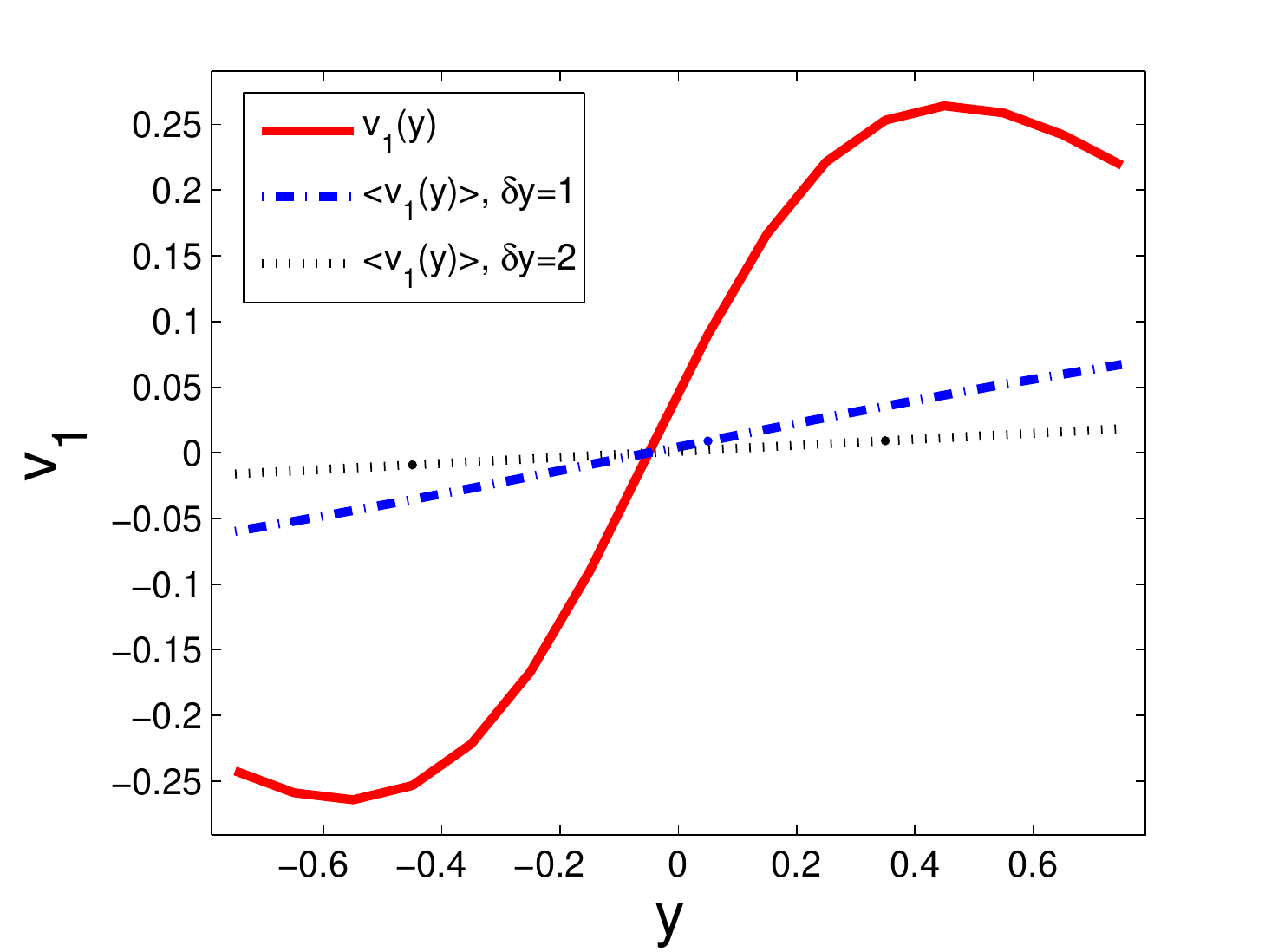}
\vskip -5mm
\caption{(color online) The $v_1$ parameter calculated for ideal massless pion 
Juttner gas, versus the rapidity $y$ for $b = 0.7 b_{max}$, at
$t = 8$ fm/c  FO time. Full curve presents semi analytical calculations 
according to eq. (\ref{flow_y}); the $v_1$ peak appears at positive
rapidity, in contrast to lower energy calculations and measurements. 
The dash-dotted and dotted curves present $v_1$ calculated taking 
into account initial CM rapidity fluctuations. }
\vskip -6mm
 \label{Fig5-v1y}
\end{figure}

{\bf Results: rapidity-dependence of the flow.\ \ }
The  $v_1(y)$ dependence 
is shown in Fig. \ref{Fig5-v1y}  (full line is the analytic solution, 
eq. (\ref{flow_y})). 
As we can see the $v_1$ is relatively large in the 
experimental rapidity range $|y|\le 0.8$, reaching a peak 
of 26 \% at $y=\pm 0.5$.  The
most important change with respect to the similar simulations for RHIC 
\cite{Baeu07}
 is that the $v_1$ now peaks in "forward" direction, 
i.e. the positive peak appears now at positive rapidity.

Qualitatively our results agree with the simulations performed in a microscopic transport model, namely the quark gluon string model \cite{v1_cascade}, where $v_1(\eta)$ in "forward" direction was obtained. However, the authors of \cite{v1_cascade} have not found the reason for such a behaviour, and have qualitatively attributed it to the different viscosities in the region with $|\eta|<3$ and at higher preudorapidities.  

\vspace{-4mm}

At lower energies in the same FD model calculations we obtained 
the $v_1$ peaking in the "backward" direction 
({\it 3rd flow component}) \cite{CsR99,Baeu07},
of a magnitude of $5$ \% and $2-3$ \% for 158 and 65 + 65 
A$\cdot$GeV energy respectively.
The position of the peaks also moved
from $|y|\approx 1.5$ to  $|y|\approx 0.5$ with energy increasing from SPS
to RHIC.  Experimentally the 3rd-flow component
was indeed measured at these energies \cite{CsR99,RHIC_V1,Wang07}, 
although the peak values were smaller. 
Especially at the RHIC energies \cite{RHIC_V1}, where the highest values were 
$v_1 \approx 0.6$ \% and $0.2$ \% for  for 62.4 + 62.4 and 200 
A$\cdot$GeV energy respectively. The peaks appeared at $|y|\approx 1$
around the far end of the acceptance of the central TPC. Thus, at RHIC 
the $v_1$ magnitude was about 5 times smaller
than the FD prediction. Also, the move towards the more central
rapidities was weaker in the experiment than in FD calculations.

The reason for such a disagreement is the 
effect of  initial state Center of Mass (CM) rapidity fluctuations, which may be decisive in the case of
$v_1$, due to the sharp change around $y=0$. 

{\bf Initial state CM rapidity fluctuations.\ \ }
One  has to take into account that the 
CM rapidity is not exactly the
same for all collisions, due to random fluctuations in the 
initial state, where the numbers of  participant nucleons from projectile
and target may not be exactly the same. This leads to 
considerable $y_{CM}$ fluctuations at large impact parameters, where the 
flow asymmetry is the strongest, while total number of participants
is the smallest.  Although, several initial state
models generate EbE fluctuating initial states, see for example \cite{v1_fluct}, longitudinal fluctuations
are not analysed up to now, neither theoretically nor experimentally in detail.
A high acceptance experiment could provide a good estimate for the EbE 
initial state rapidity, $y_{CM}$ \cite{Dieter}, which is a 
conserved quantity, i.e. it can not be changes by the system expansion, hadronization or freeze out. 

To analyze the consequences of these fluctuations, 
we assumed a Gaussian $y_{CM}$ distribution, centered at $y_{CM}=0$, 
with variance $\delta y= 1, 2$.  

\begin{figure}[h]
 \centering
 \includegraphics[width=3.3in]{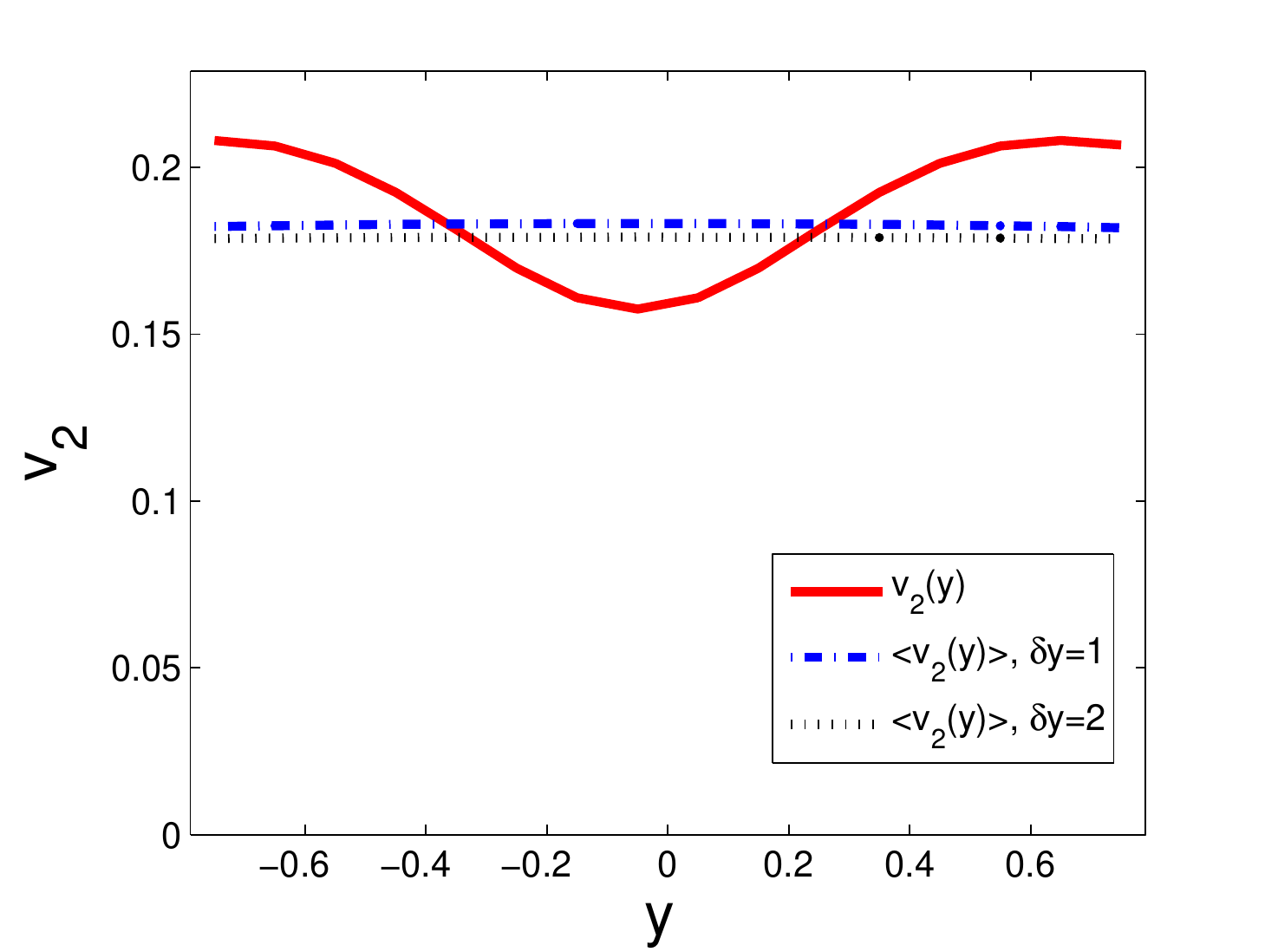}
\vskip -5mm
 \caption{(color online) 
The $v_2$ parameter calculated for ideal massless pion Juttner gas,
versus the rapidity $y$ for $b = 0.7 b_{max}$, at
$t = 8$ fm/c  FO time. Full curve presents semi analytical calculations 
according to eq. (\ref{flow_y}); dash-dotted and dotted curves present
$v_2$ calculated taking into account initial CM rapidity fluctuations.}
\vspace{-6mm}
\label{Fig6-v2y}
\end{figure}

Results can be seen  in Fig. \ref{Fig5-v1y}: dash-dotted and dotted 
lines. As expected the initial state fluctuations  strongly 
reduce $v_1(y)$ at central rapidities. 
The resulting $v_1$ 
is still large enough to demonstrate the "rotation effect", discussed above,  
however being of the order of $1\%$ it can be easily masked by the directed flow generated by the random fluctuations in the initial state. 

The first preliminary results from LHC \cite{QM2011} show $v_1(\eta)$ at midrapidity of the order of $0.1\% $ or less, in antiflow direction. Taking into account the error bars, the observed directed flow at LHC is very little, practically compared with $0$. This might be a result of a compensation of the $v_1(\eta)$ in "backward" (anti-flow) direction, coming from the random fluctuations in the inital state \cite{schenke,v1_fluct}, with the Global directed flow in the "forward" direction, as predicted by our simulations.

It is interesting to study the effects of the initial CM rapidity 
fluctuations on other observables. 
Fig. \ref{Fig6-v2y} shows the elliptic flow as a function of rapidity.   
Fluctuations make the $v_2(y)$ peaks wider, but the magnitude is hardly 
reduced.  Thus, we predict a plateau-like shape of the 
elliptic flow distribution.

The CM rapidity fluctuations have, in principle, no effect of the 
$y$-integrated $v_2(p_t)$ and $v_1^S(p_t)$ (therefore the full line in 
Figs. \ref{Fig4-v2pt}, \ref{Fig6-v1spt-b5b7}  are marked as 
"analytical solution (+ fluctuations)").  
However, in the realistic simulations, we should not integrate
over $y$ from $-\infty$ to $+\infty$, but only over the measured rapidity
range, i.e. $-0.8 \le y \le  0.8$. 
Such a "limited range" effect is 
dramatic for  $v_1^S(p_t)$ which can be reduced to less than 1\%, 
see the dashed and dotted lines in Fig. 
\ref{Fig6-v1spt-b5b7}. The $v_2(p_t)$ dependence is weakly affected
(Fig. \ref{Fig4-v2pt} dashed line).

Interestingly, the initial $y_{CM}$-fluctuations lead to some 
increase of the elliptic flow, $v_2(p_t)$, putting it in a reasonable 
agreement with the ALICE data \cite{ALICE-Flow1}, see Fig. \ref{Fig4-v2pt}, 
- please note that no fine-tuning was done. At the same time 
$y_{CM}$-fluctuations strongly reduce $v_1^S(p_t)$, see  
Fig. \ref{Fig6-v1spt-b5b7}.

{\bf To summarize,\ \ } our FD simulations of the LHC heavy ion collisions suggest that collective directed
$v_1(y)$ flow and newly introduced $v_1^S(p_t)$ function can and should 
be measured \cite{Dieter}, although these are strongly suppressed 
due to initial state $y_{CM}$-fluctuations (see Figs. \ref{Fig6-v1spt-b5b7},  \ref{Fig6-v2y}). 
For the first time in hydrodynamical calculations we see is that the $v_1$ Global flow
can change the  direction to "forward" in contrast to what happened at lower energies.
This is a result of our tilted and moving initial state \cite{MCs001}, 
in which the effective "angular momentum"  from the increasing
beam momentum is superseding the expansion driven by the pressure.
We have also proposed a new method to distinguish contributions to $v_1(p_t)$ 
from Global flow (i) and from random fluctuations in the initial state (ii). The method is based on  
$v_{1}^S(p_t)$ function, introduced by us in this work, and consist in analyzing its even and odd components
according to eqs. (\ref{v1s_sep}).

%
This work is supported in part by the European Community-Research
Infrastructure Integrating Activity HadronPhysics2, Grant \# 227431
under the 7th Framework Programme. V.K.M. also acknowledges support
from  MICINN, Spain (contract FIS2008-01661) and 
Ge\-ne\-ra\-li\-tat de Catalunya (contract 2009SGR-1289).


%


\begin{thebibliography}{17}%

\bibitem{ALICE-Flow1}%
  K.~Aamodt {\it et al.}  [The ALICE Collaboration],
  Phys. Rev. Lett. 105, 252302 (2010).

\bibitem{QM2011}
R. Snellings,  presentation at the 22nd international Conference on Ultra-
Relativistic Nucleus-Nucleus Collisions (Quark Matter 2011), Annecy, France, May 23-28, 2011; 
I. Shelyuzenkov,  {\it Ibid.}; G. Eyyubova,  {\it Ibid.}


\bibitem{RHIC_V1}
  B.~I.~Abelev {\it et al.}  [STAR Collaboration],
  Phys.\ Rev.\ Lett.\  {\bf 101}, 252301 (2008).

\bibitem{PP_method_QM}
  N.~Borghini, P.~M.~Dinh and J.~Y.~Ollitrault,
  Phys.\ Rev.\  C {\bf 64}, 054901 (2001);
  J. Schukraft,  presentation at the 22nd international Conference on Ultra-
Relativistic Nucleus-Nucleus Collisions (Quark Matter 2011), Annecy, France, May 23-28, 2011.

\bibitem{schenke}
B. Schenke, presentation at the 22nd international Conference on Ultra-
Relativistic Nucleus-Nucleus Collisions (Quark Matter 2011), Annecy, France, May 23-28, 2011.

\bibitem{Ko}
  L.W. Chen, C.M. Ko,   J. Phys. G {\bf 31}, S49 (2005).

\bibitem{v1_cascade}
  J.~Bleibel, G.~Burau and C.~Fuchs,
  Phys.\ Lett.\  B {\bf 659}, 520 (2008).


\bibitem{coalesce}%
  D. Moln{\'a}r and S.A. Voloshin,
  Phys. Rev. Lett. {\bf 91}, 092301 (2003); V. Greco, C.-M. Ko and P. L{\'e}vai,
  Phys. Rev. C {\bf 68}, 034904 (2003); R.J. Fries, B. M{\"u}ller, C. Nonaka and S.
  A. Bass, Phys. Rev. C {\bf 68}, 044902 (2003).

\bibitem{Molnar}%
  D. Moln{\'a}r and M. Gyulassy, Nucl.
  Phys. A {\bf 697}, 495 (2002), {\it erratum-ibid} A {\bf 703}, 893 (2002).

\bibitem{Son}%
  P.K. Kovtun, D.T. Son and A.O.
  Starinets, Phys. Rev. Lett. {\bf 94}, 111601 (2005); 

\bibitem{CKM}%
  L.P.~Csernai, J.I.~Kapusta,
  L.D.~{McLerran}, Phys.~Rev.~Lett.~{\bf 97}, 152303 (2006).

\bibitem{MCs001}%
  V.K.~Magas, L.P.~Csernai, and D.D. Strottman,
  Phys. Rev. C {\bf 64} (2001) 014901; Nucl. Phys. A {\bf 712}, 167 (2002).

\bibitem{CC10}%
  L.P.~Csernai, Y.~Cheng, V.K.~Magas,
  I.N.~Mishustin, D.~Strottman, Nucl.~Phys.~A {\bf 834}, 261c (2010).

\bibitem{CC09}%
  L.P.~Csernai, et al., 
  J.~Phys.~G {\bf 36}, 064032 (2009).

\bibitem{BC98}%
  L.~Bravina, L.P.~Csernai, P.~L{\'e}vai, and D.~Strottman, 
  Phys.~Rev.~C~{\bf 50}, 2161 (1994).

\bibitem{HSz}%
  Sz. Horvat, V.K. Magas, D.D. Strottman, and L.P. Csernai, 
  Phys. Lett. B {\bf 692}, 277 (2010).

\bibitem{CsR99}%
  L.P. Csernai, D. R{\"o}hrich, Phys.
  Lett. B {\bf 458}, 454 (1999).

\bibitem{Baeu07}%
  B. B{\"a}uchle, et. al., 
  J. Phys. G {\bf 34}, s1077 (2007).


\bibitem{Yun10}%
  Yun Cheng, et al., Phys. Rev. C {\bf  81}, 064910 (2010).


\bibitem{DO85}%
  P. Danielevicz, G. Odyniecz, Phys.
  Lett. B {\bf 157}, 146 (1985).


\bibitem{Wang07}%
  G. Wang, et al., (STAR Coll.), J.
  Phys. G {\bf 34}, s1093 (2007).

\bibitem{v1_fluct}
  F.~G.~Gardim, F.~Grassi, Y.~Hama, M.~Luzum and J.~Y.~Ollitrault,
  arXiv:1103.4605 [nucl-th].

\bibitem{Dieter}%
  Dieter R\"ohrich (ALICE Coll.), private communication. 

\end{thebibliography}
\end{document}